**Title:** A low-cost LLM method to extract lifespan curves from published articles

**Running title:** LLM extraction of lifespan curves

**Author:** Edouard Debonneuil, PhD (ORCID: 0009-0003-2483-9986)

**Affiliation:** International Longevity Alliance, Sceaux, France

**Corresponding author:** Edouard Debonneuil, edebonneuil@yahoo.fr

## Abstract

Science builds on the results of earlier studies. In aging research this is hard: the vast majority of survival plots come in widely varying formats and were never extracted, so past lifespan results are known only in part, as a vague impression, and cannot be pooled or re-analysed. We present a method that reads all survival curves of a figure with a single prompt to a low-cost vision-language model. Each figure is read twice, the reading that lies closer to the figure is kept, and an automatic check flags doubtful figures for manual review. On synthetic Kaplan–Meier plots, its accuracy matched that of a recent agent-based method at about a thousandth of its cost. On published lifespan figures of mice, voles, flies and worms, compared with deposited individual data, the extracted curves deviated from the truth by about 2% of the plot area on average. On a random sample of the mammalian aging literature published between 1961 and 2021, processed after the method had been developed on another sample, 97% of survival figures were read correctly, and the 3% clear failures were all among the 29% of figures flagged for manual review. Survival curves can thus be extracted from the whole aging literature for a fraction of a cent per figure, towards aging research that is better grounded in the work of past generations of researchers.

## Introduction

Survival curves are the primary result of most lifespan experiments, yet the individual death ages behind them are rarely deposited. Comparing interventions across studies, species and decades therefore requires reading numbers off published figures. Meta-analyses of dietary restriction in rodents, for example, relied on summary statistics extracted by hand from heterogeneous reports (1). Programs that release individual data, such as the NIA Interventions Testing Program (ITP) through the Mouse Phenome Database (2,3), remain the exception. Curated databases such as DrugAge (4) and the Rodent Aging Interventions Database (RAID) of the LEV Foundation (5) compile the effects of interventions on lifespan as summary values (for example, the change in median or maximum lifespan), not the full survival curves.

Digitising curves is a well-studied problem in clinical research, where reconstructed individual patient data feed meta-analyses (6,7). Point-and-click tools, with which a person calibrates the axes and clicks along each curve, are accurate but slow. SurvdigitizeR automates digitisation with classical image processing, but needs the axis ranges and the number of curves for each figure (8). Recent work uses large language models: KM-GPT combines image processing with a frontier multimodal model (GPT-5) called for each plot, and is offered as a web interface rather than as code (9); KMGen lets a model write and iterate image-processing code until the curve overlays match (10), which with Claude Opus costs a few dollars per plot (see Results). These systems are precise, but target clinical trial figures with numbers-at-risk tables and well-defined Kaplan–Meier steps, and their cost or mode of access does not suit the batch processing of a whole literature.

Lifespan figures in aging biology come in far more varied formats. They span six decades of plotting software, use percentages or numbers of animals alive, often show markers or connected points rather than steps, follow cohorts from birth or from an advanced age, and are frequently assembled into multi-panel figures with shared legends. They are also numerous: lifespan experiments number in the tens of thousands in rodents alone, and many more in flies, worms and other species screened at scale. Until recently, extracting their results seemed to require human hands. In 2023, with the Longévité & Santé association and the International Longevity Alliance, the author launched COLLIDA (Collective Collection of Lifespan Data), a volunteer project to collect the lifespan data of published experiments, so that new research against aging could build on what is already known rather than proceed blind to it; the diversity of formats made manual collection a daunting task, and the project was set aside as language models able to read texts and figures emerged. Its collection of more than 2,000 articles is the corpus used here.

Here we present and evaluate a low-cost method for this task. Its core is a single prompt to a general vision-language model, with no code generation; the model reads the axes and the number of curves itself, so that no manual input is needed. The method handles the formats of aging research (points, numbers of animals alive, late-starting cohorts, multi-panel figures), runs in batch, and flags doubtful readings for a human check. We evaluate it on synthetic plots built with the KMGen protocol, on real published figures whose ground truth we rebuilt from deposited individual lifespans, and on a random sample of the aging literature.

## Methods

### The method

The method (Figure 1) has one core step and three complements.

**Core: a single prompt.** All images of survival figures are sent with one prompt, in calls of up to three images (figures with four or more curves are sent alone). For each graph the model returns its type, decided from how the curves are drawn (Kaplan–Meier staircase, or values joined by sloping segments or a smooth line), both axes (quantity, unit, first and last labelled tick, tick step) and, for each curve named as in the legend, either the list of vertical drops (time, survival after the drop) and censoring marks, or, for other plots, a list of points chosen according to curvature. For each curve it also returns where the line begins and ends, so that a cohort followed from an advanced age starts at that age and a final plateau after the last drop is kept; a curve with no visible death is returned as a flat line rather than omitted. Asking for the drops of a staircase rather than for points along a curve is the main design choice of the prompt. The response format is imposed through the structured-output option of the API (a JSON schema), so that every response can be read by a program. Images narrower than 800 pixels are enlarged (Lanczos) before the call. Values are converted to survival fractions with one scale per graph, never point by point, so that a curve falling to 1% is not read as 100%: percentages are divided by 100 (unless the axis is printed from 0 to 1), and numbers of animals alive by the starting count of each curve. Times read as all negative on a time axis (labels printed with a leading dash) are turned back positive, and each survival curve is replaced by the closest non-increasing curve (isotonic regression). The model is gpt-5.6-luna (OpenAI), used through the Responses API with low reasoning effort; preliminary tests by the author with Qwen3.6-Flash, Qwen3.8-Flash, Gemini 3.5 Flash-Lite, Claude Haiku 4.5 and Claude Sonnet 5 did not reproduce test curves as closely (not reported). The full prompt and schema are given in the Supplementary Material.

**Complement 1: SurvdigitizeR for precision.** For graphs classified as Kaplan–Meier, SurvdigitizeR (8) is run with the number of curves and the axis ticks read by the model, which removes its manual input. Its curves are matched to the model's curves and kept only if the two readings agree (mean IAE between them below 0.02); otherwise the model's curves are kept.

**Complement 2: panel splitting.** For figures with several panels, a short additional call returns the bounding box of each survival panel, of a legend located in another panel, and of an x-axis strip shared with another panel. Each panel is then sent as its own image, with the shared legend and the x-axis strip (cropped over the same horizontal range, so that ticks stay aligned) pasted underneath. A split is rejected, and the figure kept whole, if the panels cover less than half of the figure or if a panel is smaller than 40% of the figure in both dimensions.

**Complement 3: two readings, one kept, doubtful figures flagged.** Each figure is read twice with the same prompt. Both readings are drawn back onto the image, the axes being located from the printed tick labels (read with Tesseract), the tick marks or the axis lines, and the reading whose points lie closer to the ink of the figure is kept (mean share of points within 0.5%, 1%, 2% and 4% of the image size of a dark pixel). The figure is then flagged for manual checking when (i) the two readings disagree (a curve missing in one of them, or two readings of a same-named curve differing by an IAE above 0.05, about three times the typical error); (ii) a curve stops before 80% of the time reached by the longest curve while still above 5% survival, typically a misread end; (iii) fewer than 70% of the points of the reading kept land on ink; or (iv) the figure is unlikely to show survival (time not on the x axis, all curves rising, or a y axis that is not survival). Apart from the second reading, these steps use no model call.

In the full COLLIDA pipeline these steps are preceded by text extraction (with OCR for scanned articles), a text-only call that identifies survival figures and their experimental groups, and a vision call that crops each figure from the page.

## Synthetic benchmark

The images and ground truth of the KMGen benchmark (10) are not public, so we regenerated 32 plots following its published protocol: Weibull event times with random censoring, exact Kaplan–Meier step curves drawn with matplotlib, in four categories (5 standard plots, 7 edge cases, 12 single degradations and 8 combined degradations such as blur, JPEG compression, low resolution, grids, annotations, black-and-white rendering, many arms and multiple panels). Because the random draws differ, comparisons with the published KMGen figures are indicative only.

## Real figures with known individual data

**ITP.** Individual lifespans of the 2009 ITP cohort (2,332 mice) were obtained from the Mouse Phenome Database (3). We checked that the counts and the sum of ages matched the original file, and that the recomputed median lifespans matched those published (e.g., males: control 807 days, rapamycin 4.7/14/42 ppm 834/909/992 days) (11). Figures came from the PubMed Central open-access collection: Figures 1A, 1B, 5A, 5E and 5F of Harrison et al. 2014 (acarbose, 17-α-estradiol, methylene blue) (12) and both panels of Figure 1 of Miller et al. 2014 (rapamycin) (11), giving 7 panels and 16 curves. Ground truth was the Kaplan–Meier estimate pooled across sites.

**Dryad.** We searched the Dryad repository for lifespan datasets (344 hits) and kept those with an open-access article in PubMed Central whose survival panel could be matched unambiguously to deposited individual data. Seven studies were retained (13–19): flies (insulin-cell ablation × sugar; Toll-pathway RNAi), voles (vitamin supplementation), worms (daf-2 RNAi; temporary fasting; dietary restriction regimens) and mice (Cdc42 inhibition). They contribute 14 images and 60 curves, drawn with base R, ggplot2, Excel, Prism and other software, as steps, lines with markers and dashed lines. Six images contain two panels (side by side or stacked) whose legend or x axis is shared. Ground truth was the Kaplan–Meier estimate from individual ages, with the censoring rule of the original authors (for example, worms dead by matricide were censored (19)); for one study the deposited spreadsheet gives the plotted percent-survival points, which were used directly. Every ground-truth curve was overlaid on the published figure before scoring.

## A random sample of the literature

To see how the method behaves on the literature it is meant for, including old scanned articles without deposited data, the method was developed on a random sample of 129 articles of the COLLIDA collection (2,315 articles, 1940–2022) and then tested on a second random sample of 62 other articles, drawn with a fixed seed as one article per publication year (years before 1960 grouped by decade) and processed with the full pipeline. The rule for keeping one of the two readings, chosen on the benchmarks, and the exception for percent axes printed from 0 to 1 were added after the test sample had been processed; all other steps were fixed before it was drawn. We kept the survival figures of mammals that were correctly cropped and counted crop failures separately. Without individual data, every reading was checked visually against the published figure, the extracted curves being drawn back onto it and also re-plotted with the axis ranges read by the model, and classified as correct, imprecise or incomplete, clear failure (a curve following the wrong line, missing or swapped), or unverifiable. Each figure was first classified by Claude (Anthropic) and the classification was then checked by the author, whose verdict was kept; the review files are available with the code (see Data Availability).

## Metrics, statistics and cost

We used the KMGen metrics (10). The IAE is the integral, over time normalised to the x-axis maximum, of the absolute difference between extracted and true survival, evaluated with left-continuous step interpolation (linear interpolation for point curves) on the merged time grid. We call a clear failure an arm with IAE above 0.10 against the truth, or an arm that could not be matched by name. Arms were matched to the truth by name only (Hungarian assignment on string similarity), never by shape; unmatched arms were counted, not scored. Each benchmark was read three times; the reading kept was chosen within each pair of repetitions, and per-image IAE (mean over arms, then over repetitions or pairs) was compared with paired Wilcoxon signed-rank tests. Costs are given at list prices (gpt-5.6-luna: US$0.20 and $1.20 per million input and output tokens; Claude Opus 4.6: $5 and $25), from the API usage records. To measure the cost of the original

KMGen, its instructions were run unchanged with Claude Code on Claude Opus 4.6, as in the original study, on five synthetic plots covering the four categories.

### Use of AI tools

The method under evaluation is itself an AI system (gpt-5.6-luna, OpenAI). In addition, Claude (Anthropic; Claude Code) was used to write the benchmark and analysis code, to retrieve and assemble the public datasets, to run the experiments, to make a first classification of the figures of the literature sample, and to draft the manuscript. The author designed the study and the method, checked the code, the ground-truth overlays and all results, revised the text, and takes full responsibility for the content.

## Results

### Synthetic plots

The method reached a mean IAE of 0.015 (median 0.014) over the 32 plots, the same mean as published for KMGen (Table 1). Accuracy hardly depended on image quality: blur, JPEG compression, grids, annotations and black-and-white rendering left the error close to that of clean plots (Figure 3). On clean plots, the error (0.010) was twice that of KMGen (0.005), because the model summarises each staircase with about 20 drops where the true curve has about 90 steps; crowded figures with four to eight arms were less precise, and a tiny blurred plot produced the only clear failures (2 of 237 arm readings, in one repetition).

SurvdigitizeR contributed most of the gain on clean plots. Used alone, given the axes, it returned no output for 9 of 32 plots (28%), was often very precise when it worked (IAE 0.005–0.015 on 14 plots), but on the other 9 was grossly wrong without warning (0.04–0.40): touching curves were merged, black-and-white curves could not be separated, and annotations were read as curves. Within the method, with the axes and number of curves read by the model, it was retained for 32% of plots and lowered the mean IAE from 0.016 to 0.015 (better on 9 plots, worse on 3, $p = 0.01$); no large SurvdigitizeR error passed the agreement test (worst retained IAE 0.015). Keeping the closer of two readings made no difference on synthetic plots, whose two readings rarely differed.

## Real figures

On the seven ITP panels, the model identified every arm and returned valid output in all calls. The mean IAE was 0.016 (median 0.015), close to the synthetic benchmark (Table 2, Figure 2); 1 of 54 arm readings was a clear failure, on the four closely intertwined rapamycin curves in males. On the 14 Dryad images, the mean IAE was 0.022 (median 0.021), whatever the organism or plotting software: 0.019 on single-panel and 0.025 on two-panel images. Keeping the reading closer to the figure, rather than the first reading, lowered the mean error from 0.017 to 0.016 on ITP panels and from 0.028 to 0.022 on Dryad images, close to what a perfect choice between the two readings would give (0.015 and 0.020; differences not significant on these small sets). Panel splitting mattered: during development, without it, the mean IAE on two-panel images was 0.054 against 0.036 with it (paired $p < 0.001$), as the model mixed panels or merged their curve names. Its cost is identity: when a legend placed in a neighbouring panel was not re-attached, curves were named by colour ("red solid"), and 10 of 180 arm readings could not be matched by name. SurvdigitizeR was never retained on real figures: it could not locate the axis ticks of six of the seven ITP panels and was grossly wrong on the seventh (IAE 0.19), and it produced an output for only 7% of Dryad runs.

## Cost

One reading cost about US$0.001–0.003 per figure (about 1,000 input and 1,000 output tokens). The original KMGen agent on Claude Opus 4.6 needed 14 to 57 turns, 3 to 37 minutes and $0.98–5.93 per plot (mean $2.88), that is 365 to 4,650 times more than the single prompt (median 1,150-fold); in exchange it was more precise on these five plots (mean IAE 0.005, range 0.002–0.013, against 0.025, range 0.008–0.075). Processing whole articles, figure identification and two readings included, cost about 1.5 cents per article (US$0.90 for the 62 articles of the test sample).

## A random sample of the literature

The 62 articles of the test sample yielded 134 figures identified as survival figures, of which 23 were not (dose–response, life expectancy or body weight plots), 8 were badly cropped (a panel cut or several graphs left in one crop) and 4 could not be verified (small figures whose groups could not be told apart); the detector flagged all 35. The other 99 survival figures of mammals came from 42 articles published between 1961 and 2021, 30 of them from 15 articles published before 1990, many scanned. Of these, 96 (97%) were read correctly and 3 (3%) were clear failures: in one figure a curve of four was kept at 100% until its last death, in a figure without legend one of two curves was missed, and in a small 1961 cumulative-mortality figure one curve of six was missed. The detector flagged all three, together with 26 of the 96 correct figures (27%), so that 29 figures (29%) were flagged in all; none of the 70 unflagged figures had a problem. In one figure of a cohort followed from an advanced age, one of the two readings started the curves at birth; the reading kept was correct. Reading the flagged figures again with high instead of low reasoning effort brought no clear gain.

## Discussion

A single prompt to a low-cost vision-language model, without code generation, recovered published survival curves with a mean integrated error of about 2% of the plotted area on real figures from mice, voles, flies and worms. With SurvdigitizeR as a complement, its mean error on synthetic plots matched that of KMGen, an agent that writes and verifies its own image-processing code (10), at about a thousandth of its cost; KMGen remains more precise plot by plot. On a random sample of the literature spanning six decades, including old scanned figures, 97% of survival figures were read correctly, and the clear failures were all among the figures flagged for a manual check. These errors are small compared with the lifespan effects that most intervention studies are designed to detect, and the method needs no manual input of axes or curve counts.

This makes it realistic to extract the survival curves of the whole lifespan literature. Today, research on aging advances largely blind to what generations of researchers obtained before: only part of their results is known, mostly as a vague impression of which interventions extended lifespan, in which species and by how much, rather than as data that can be pooled and re-analysed. At a fraction of a cent per figure, with about a third of figures to check by hand, the tens of thousands of published lifespan experiments can become such data: survival curves from which individual lifespans can be reconstructed (6,7), compared across studies, strains and decades, and re-analysed with current methods.

Three findings are, in our view, more general than the specific model tested. First, errors depend much more on figure layout than on image quality or plotting style: crowded figures and multi-panel images drive most of the error, and handling the layout (one panel per image, shared axes and legends re-attached) matters more than the wording of the prompt. Second, classical digitisation software, accurate on clean synthetic plots, fails silently or completely on real lifespan figures; it is worth keeping only when an independent reading agrees with it. Third, ground truth rebuilt from deposited data must itself be checked: a crop or a censoring rule different from the authors' is visible at a glance on an overlay of the truth on the published figure, and would otherwise be counted as an error of the method. We recommend that digitisation benchmarks publish such overlays.

This study has limitations. The real-figure sets with known individual data are small (21 images, 76 curves) and limited to figures with deposited data, which may be cleaner and more recent than the literature as a whole. The literature sample extends the evaluation to older and scanned figures, but was checked visually rather than against individual data, and it is small: with no clear failure among 70 unflagged figures, the rate of unflagged failures could still be a few percent. Two steps, the choice of the reading kept and the exception for 0–1 percent axes, were added after the test sample had been processed. Choosing the reading closer to the ink corrects shifted or misplaced readings, but cannot tell which of two neighbouring lines a curve should follow, nor notice a missing curve. The upstream steps (figure identification and cropping) were not evaluated in detail, although the figures they got wrong in the test sample were all flagged. The synthetic comparison with KMGen uses regenerated plots, and the cost of the original KMGen was measured on five plots only. The model is proprietary and may change or be withdrawn; we therefore release the prompt, raw responses and scores so that other models can be compared on the same material. Several improvements are within reach: giving the model the figure caption, which names curves when the legend is missing, and the age at entry stated in the text; choosing the better reading curve by curve rather than for the whole figure; and, for flagged figures, a light agent in the spirit of KMGen run on the same low-cost model, which in our tests reduced some errors but was not reliable enough to replace a human check. Next steps are to digitise by hand a sample of older scanned figures, and to reconstruct individual lifespans from the extracted curves (6,7) for meta-analysis.

## Funding

This work was funded by Longevity Group SAS (Paris, France).

## Conflict of Interest

E.D. is a paid consultant for Longevity Group SAS, which funded this work. The study was conceived by the author before this collaboration. The funder asked for the work to be submitted for publication but had no role in the study design, data collection, analysis or interpretation, or in the writing of the manuscript.

## Data Availability

Code (pipeline and benchmark scripts), prompts, synthetic images, ground-truth curves, raw model responses and scores are available from the corresponding author on request and will be deposited in a public repository (GitHub, archived on Zenodo) upon acceptance. Individual lifespan data are available from the Mouse Phenome Database (ITP C2009 cohort) (3) and from Dryad (datasets cited in references 13–19); published figures are available from PubMed Central.

## Author Contributions

E.D. designed the study and the method, checked the code and all results, and wrote the manuscript (see Use of AI tools).

## Acknowledgments

We thank the volunteers who annotated articles for the COLLIDA project: Malvina Pushkova, Dave Ringelberg, Anton Kulada, Leon Peshkin, Walter Crompton, Alexandra Stolzing, Patrick Wilson, Josiah Akinloye, Benyamin Matei-Dediu, François Robin-Champigneul, Olga Spiridonova, Laure Pouliquen, Peter-Mikhael Richard, Alex Sverlov, Abasse Dabere, Sven Bulterijs, Martin O'Dea, Alexey Moskalev, Aaron King, Richard Herrell, Morand Studer, Thomas Hahn, Jean Allemand, Harold Knoll, Augustin Terlinden, Francesco Neri, and the other members and supporters of Longévité & Santé and the International Longevity Alliance who took part. We also thank the authors of the cited studies and the Interventions Testing Program for depositing individual lifespan data. Generative AI was used as described in Methods (Use of AI tools): gpt-5.6-luna (OpenAI) is the method evaluated, and Claude (Anthropic; Claude Code with Claude Opus models) was used to write the analysis code, run the experiments, make a first classification of the figures and draft the manuscript, under the direction and with the review of the author.

## References

1. Swindell WR. Dietary restriction in rats and mice: a meta-analysis and review of the evidence for genotype-dependent effects on lifespan. *Ageing Res Rev.* 2012;11:254–270. doi:10.1016/j.arr.2011.12.006

2. Nadon NL, Strong R, Miller RA, et al. Design of aging intervention studies: the NIA interventions testing program. *Age (Dordr).* 2008;30:187–199. doi:10.1007/s11357-008-9048-1

3. Bogue MA, Ball RL, Walton DO, et al. Mouse Phenome Database: curated data repository with interactive multi-population and multi-trait analyses. *Mamm Genome.* 2023;34:509–519. doi:10.1007/s00335-023-10014-3

4. Barardo D, Thornton D, Thoppil H, et al. The DrugAge database of aging-related drugs. *Aging Cell.* 2017;16:594–597. doi:10.1111/acel.12585

5. Peto MV, Floyd AJ, Zealley B, de Grey ADNJ. The rodent aging interventions database (RAID): a data visualization tool for all studies reporting rodent lifespan extension. *Aging (Albany NY).* 2025;17:851–862. doi:10.18632/aging.206228

6. Guyot P, Ades AE, Ouwens MJ, Welton NJ. Enhanced secondary analysis of survival data: reconstructing the data from published Kaplan-Meier survival curves. *BMC Med Res Methodol.* 2012;12:9. doi:10.1186/1471-2288-12-9

7. Liu N, Zhou Y, Lee JJ. IPDfromKM: reconstruct individual patient data from published Kaplan-Meier survival curves. *BMC Med Res Methodol.* 2021;21:111. doi:10.1186/s12874-021-01308-8

8. Zhang JZ, Rios JD, Pechlivanoglou T, et al. SurvdigitizeR: an algorithm for automated survival curve digitization. *BMC Med Res Methodol.* 2024;24:147. doi:10.1186/s12874-024-02273-8

9. Zhao Y, Sun H, Ding Y, Xu Y. KM-GPT: an automated pipeline for reconstructing individual patient data from Kaplan-Meier plots. *arXiv.* 2025:2509.18141.

10. Jiang J, Gao C, Rasmussen E, Xie SZ, Sun J. KMGen: a skill-based approach for synthetic individual patient data generation. *Proc Mach Learn Res.* 2026;340:1–60.

11. Miller RA, Harrison DE, Astle CM, et al. Rapamycin-mediated lifespan increase in mice is dose and sex dependent and metabolically distinct from dietary restriction. *Aging Cell.* 2014;13:468–477. doi:10.1111/acel.12194

12. Harrison DE, Strong R, Allison DB, et al. Acarbose, 17-α-estradiol, and nordihydroguaiaretic acid extend mouse lifespan preferentially in males. *Aging Cell.* 2014;13:273–282. doi:10.1111/acel.12170

13. Al Saud SN, Summerfield AC, Alic N. Ablation of insulin-producing cells prevents obesity but not premature mortality caused by a high-sugar diet in *Drosophila*. *Proc Biol Sci.* 2015;282:20141720. doi:10.1098/rspb.2014.1720 (data: doi:10.5061/dryad.t04v4)

14. Selman C, McLaren JS, Collins AR, Duthie GG, Speakman JR. Deleterious consequences of antioxidant supplementation on lifespan in a wild-derived mammal. *Biol Lett.* 2013;9:20130432. doi:10.1098/rsbl.2013.0432 (data: doi:10.5061/dryad.31cc4)

15. Mautz BS, Lind MI, Maklakov AA. Dietary restriction improves fitness of aging parents but reduces fitness of their offspring in nematodes. *J Gerontol A Biol Sci Med Sci.* 2020;75:843–848. doi:10.1093/gerona/glz276 (data: doi:10.5061/dryad.z08kprr8k)

16. Huayta J, Crapster JP, San-Miguel A. Endogenous DAF-16 spatiotemporal activity quantitatively predicts lifespan extension induced by dietary restriction. *Commun Biol.* 2023;6:203. doi:10.1038/s42003-023-04562-2 (data: doi:10.5061/dryad.vx0k6djwt)

17. Florian MC, Leins H, Gobs M, et al. Inhibition of Cdc42 activity extends lifespan and decreases circulating inflammatory cytokines in aged female C57BL/6 mice. *Aging Cell.* 2020;19:e13208. doi:10.1111/acel.13208 (data: doi:10.5061/dryad.c4798q5)

18. Fabian DK, Garschall K, Klepsatel P, et al. Evolution of longevity improves immunity in *Drosophila*. *Evol Lett.* 2018;2:567–579. doi:10.1002/evl3.89 (data: doi:10.5061/dryad.cp38vj4)

19. Lind MI, Ravindran S, Sekajova Z, Carlsson H, Hinas A, Maklakov AA. Experimentally reduced insulin/IGF-1 signaling in adulthood extends lifespan of parents and improves Darwinian fitness of their offspring. *Evol Lett.* 2019;3:207–216. doi:10.1002/evl3.108 (data: doi:10.5061/dryad.8d2t65n)

## Tables

**Table 1.** Synthetic Kaplan–Meier benchmark (32 plots generated with the KMGen protocol; mean IAE over three repetitions, median in parentheses).

| Method | Standard (5) | Edge (7) | Single degradation (12) | Combined degradations (8) | All (32) | Cost per plot (US$) |
|---|---|---|---|---|---|---|
| Our method (single prompt + SurvdigitizeR when they agree) | 0.010 | 0.011 | 0.015 | 0.019 | 0.015 (0.014) | ≈0.002 |
| Single prompt only | 0.015 | 0.014 | 0.016 | 0.020 | 0.016 (0.015) | ≈0.002 |
| SurvdigitizeR alone, axes entered by hand* | 0.071 | 0.071 | 0.148 | 0.018 | 0.091 (0.015) | — |
| KMGen, published (other draws)† | 0.005 | 0.009 | 0.014 | 0.028 | 0.015 (0.008) | ≈2.9‡ |

*No output for 28% of plots; mean over plots with an output, curves matched to the truth by best fit, which favours SurvdigitizeR. †Values from reference 10, on plots generated with the same protocol but different random draws. ‡Measured here on five plots (range 0.98–5.93)*

**Table 2.** Real published figures with ground truth rebuilt from individual lifespans (mean IAE of the reading kept, over the pairs of three repetitions; median in parentheses).

| Set | Images | Curves | Our method | First reading only | SurvdigitizeR output |
|---|---|---|---|---|---|
| ITP 2009 cohort (mice) | 7 | 16 | 0.016 (0.015) | 0.017 | 1 of 7 panels (IAE 0.19) |
| Dryad, single-panel images | 8 | 34 | 0.019 (0.018) | | 7% of runs, never retained |
| Dryad, two-panel images | 6 | 26 | 0.025 (0.026) | | |
| Dryad, all | 14 | 60 | 0.022 (0.021) | 0.028 | |
| Unmatched arms (Dryad) | | | 10 / 180 | | |

ITP images are single panels; Dryad two-panel images are split into panels before reading (Complement 2).

## Figure Legends

**Figure 1.** The method. Core: a single prompt returns axes and named curves for each figure (blue). Complements (orange): panel splitting for multi-panel figures; SurvdigitizeR, run with the axes read by the model and kept only when it agrees with the model; a second reading, the reading closer to the ink of the figure being kept; and flags that send doubtful figures to a manual check. Grey: upstream steps of the COLLIDA pipeline (text, figure detection, cropping).

**Figure 2.** Examples of real published panels (top) and extracted curves (colour) against ground truth rebuilt from individual lifespans (black), first reading. From left: ITP rapamycin females and males (11); vole vitamin supplementation, two stacked panels sharing one x axis, after panel splitting (14); worm dietary-restriction regimens plotted as points (16). Figure panels reproduced from open-access articles published under a CC BY licence (11, 14, 16).

**Figure 3.** Synthetic plots from easy to hard (top) and extracted curves (colour) against the truth (black), single prompt, one reading. From left: a clean plot, where the model summarises each staircase with fewer, larger steps; grids, annotations and black-and-white rendering, which barely affect the error; eight crowded arms; and a tiny blurred image, the only source of clear failures on synthetic plots.

Figure 1

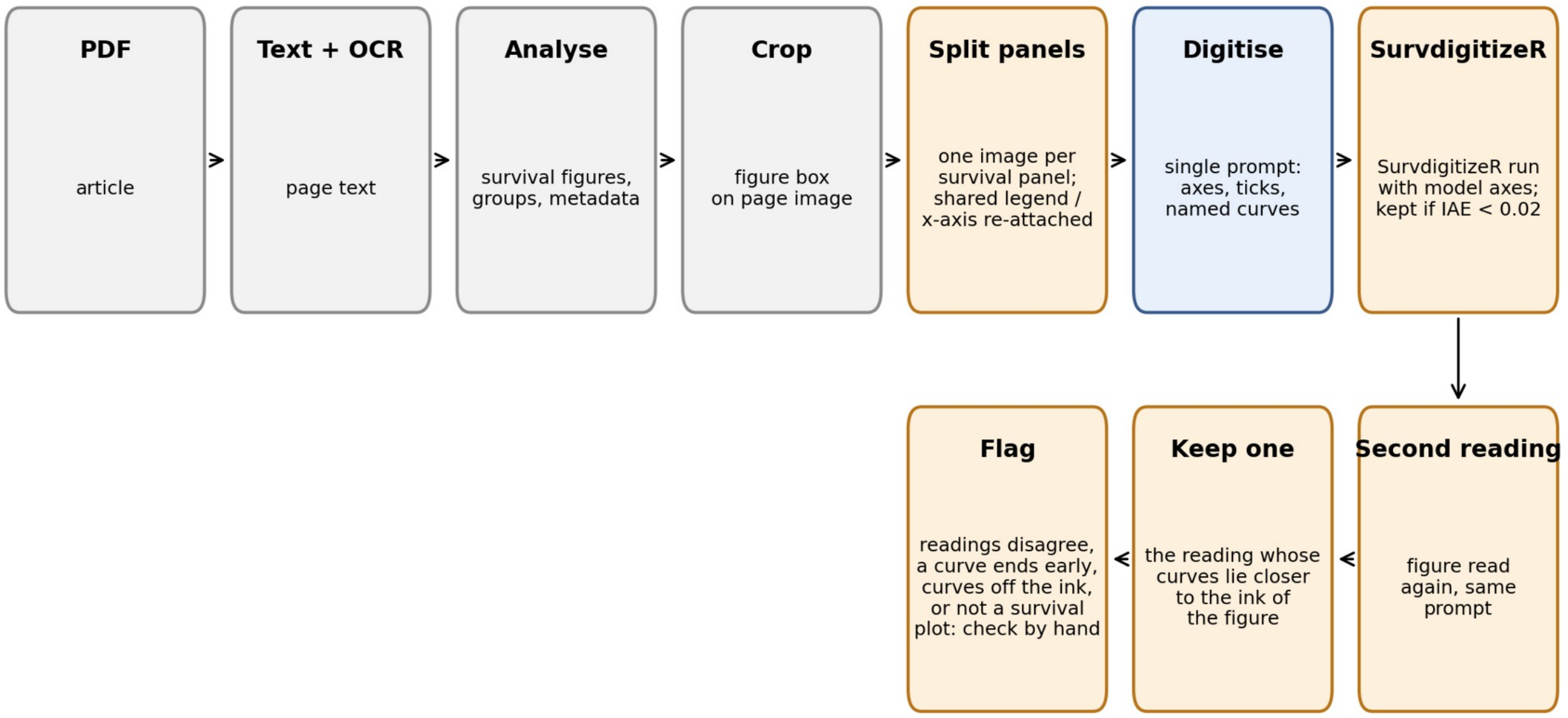

Figure 2

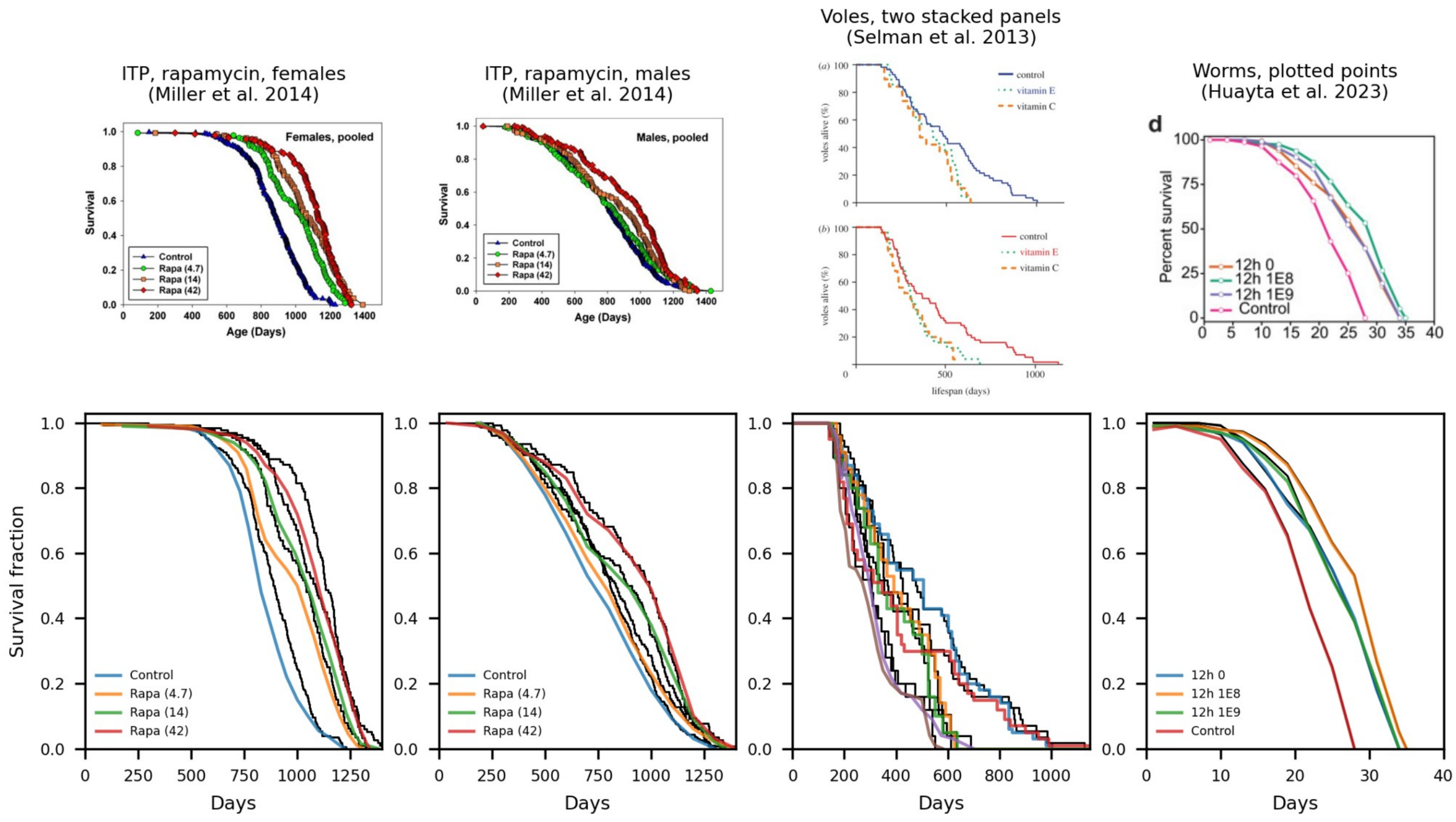

Figure 3

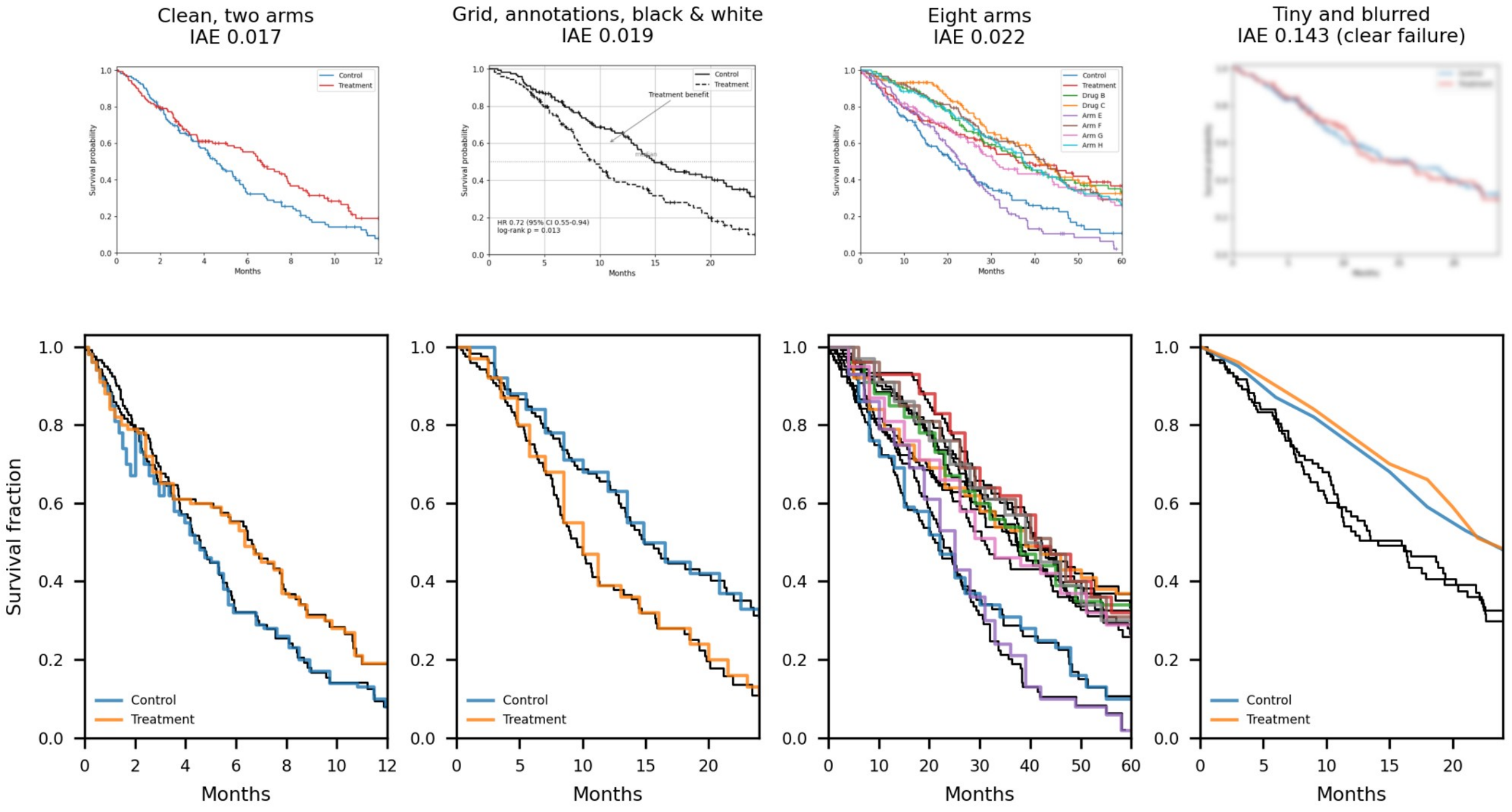
Clean, two arms
IAE 0.017
Grid, annotations, black & white
IAE 0.019
Eight arms
IAE 0.022
Tiny and blurred
IAE 0.143 (clear failure)
Survival fraction
Months
Control
Treatment